\documentclass[preprint]{JHEP3} 

\usepackage{epsfig,multicol}

\newcommand\fverb{\setbox\pippobox=\hbox\bgroup\verb}
\newcommand\fverbdo{\egroup\medskip\noindent%
			\fbox{\unhbox\pippobox}\ }
\newcommand\fverbit{\egroup\item[\fbox{\unhbox\pippobox}]}
\newbox\pippobox

\newcommand{\be}{\begin{equation}} 
\newcommand{\ee}{\end{equation}}

\newcommand{\ba}{\begin{eqnarray}}
\newcommand{\ea}{\end{eqnarray}}

\title{Level truncation and the quartic tachyon coupling}

\author{Matteo Beccaria\thanks{Partially supported by INFN, IS-LE21}\\
	Dipartimento di Fisica, Universita' di Lecce,\\
        Via Arnesano, 73100 Lecce\\
	INFN, Sezione di Lecce\\
	E-mail: \email{matteo.beccaria@le.infn.it}}

\author{Carlo Rampino\\
	Dipartimento di Fisica, Universita' di Lecce,\\
        Via Arnesano, 73100 Lecce\\
	E-mail: \email{carlo.rampino@le.infn.it}}

\preprint{\hepth{???}}

\abstract{
We discuss the convergence  of level truncation in bosonic open string field theory.
As a test case we consider the calculation of the quartic tachyon coupling $\gamma_4$.
We determine the exact contribution from states up to level $L=28$ and discuss
the $L\to\infty$ extrapolation by means of the BST algorithm. 
We determine in a self-consistent way both the coupling and the 
exponent $\omega$ of the leading correction to $\gamma_4$ at finite $L$ that we assume 
to be $\sim 1/L^\omega$. 
The results are $\gamma_4 = -1.7422006(9)$ and $|\omega-1|\lesssim 10^{-4}$.
}


\begin{document} 

\section{Introduction}
\label{Sec:Intro}

The effort dedicated to the confirmation of Sen's conjectures~\cite{Conjectures} 
has established  Witten's string field theory~\cite{OSFT} as a distinguished calculational framework 
where quantitative predictions can be readily obtained.

As is well known, Sen's conjectures concern the properties of the non perturbative vacuum of bosonic open string theory
and relate, in the simplest case, the tachyon instability to the presence of the D25-brane.
In brief, the conjectures predict that, when the tachyon condensates, a non perturbative vacuum is reached whose physics is described by 
closed string theory with no tachyon and branes at all.
The matching between the energy of the tachyon false vacuum and the D25-brane tension leads to a clear-cut prediction 
for the value of the tachyon potential at its global minimum. Therefore, the check of Sen's conjectures requires a 
reliable calculation of the potential itself. This is a typical off-shell quantity and string field theory is
an ideal framework for the calculation.

Unfortunately, an exact expression for the non perturbative vacuum  is not known at present and some kind of 
analytic approximation  is needed. In string field theory a possibility is offered by the level truncation approach
(LT) that is a systematic expansion in  string field components with increasing masses. 
The accuracy of the expansion, in a fixed gauge, is governed by a single parameter $L$ related to the Virasoro level 
of the states included in the calculation.

The application of LT to the verification of Sen's conjectures has been quite successfull up to date. Also, the same
technique has been exploited in order to collect useful numerical insights for the identification
of the correct vacuum string field theory~\cite{VSFT}.
After the initial and later revised calculations of the tachyon potential at level $L=4$~\cite{L4}, 
the check of Sen's predictions has been extended to level $L=10$~\cite{L10} and more recently up to $L=18$~\cite{L18}.
A good convergence with respect to $L$ is observed and the conjectures
are verified numerically with amazing agreement.

Some technical questions, however, remain open. Level truncation breaks the gauge invariance of Witten's action
and gauge fixing is not necessary. However it is usual to perform calculations in the Feynman-Siegel gauge 
that introduces a few simplifications (see~\cite{Gauge} for an 
exploration of other gauge choices). At the numerical level, the LT series in Feynman-Siegel gauge turned out to 
approach its asymptotic value monotonically at least in the first studies. A surprise came from the 
extended calculation in~\cite{L18} that revealed overshooting at finite $L$ in the values of the  minimum of the tachyon potential.
This observation raised some doubts concerning the best and the correct way to extrapolate the results 
to $L\to\infty$. Several strategies have been proposed based on reasonable assumptions about the 
asymptotic properties of the LT series and it appears of some interest to investigate extrapolation methods
in a more systematic way.

Actually, the problem of extrapolating short expansion series is common in the analysis of lattice models
of statistical mechanics. This is particularly true in several dimensions where, typically, only a few terms are known in the 
 available perturbative expansions ({\em e.g.} high or low temperature). Robust algorithms have been devised to deal with such cases.
This paper will be devoted to analyze the performance of one of these methods when applied to string field theory.

We shall analyze in details a specific case that is easy to 
treat with level expansion  and that permits the comparison with rather accurate alternative predictions. 
This is the quartic self-coupling $\gamma_4$ in the tachyon effective potential. 
A closed analytical expression exists~\cite{ClosedForm,Diagrams} and its evaluation can be reduced 
to numerical quadratures that are non trivial, but in principle straightforward.

The best available result is $\gamma_4 = -1.742(1)$~\cite{Diagrams}; as we shall see later, it 
has been checked with two different level truncation schemes with good agreement at least within the
numerical error.

Here, we shall first improve the analytical knowledge of $\gamma_4$ by extending the  LT series from 
the known result at level $L=20$~\cite{Taylor00} up to $L=28$. This effort involves an increase of the number of 
considered states by a factor larger than 10. Since $\gamma_4$ is certainly not the most interesting quantity to be 
studied, the motivation of this work is that we want a  long LT series to analyze
the various extrapolation algorithms in a clear way. Of course, a calculation at such level
for more complicated quantities like, for instance, the full tachyon potential would be hardly feasible
and we hope to draw conclusions holding for shorter series too.

As a second step in our analysis, we shall weaken as far as possible the 
hypotheses underlying the extrapolation procedure. In particular, we shall never assume a convergence rate with leading 
correction$\sim 1/L$, an assumption usually  made in the common approaches.
We shall apply the BST algorithm to  accelerate the  convergence of series 
with general  leading corrections $\sim 1/L^\omega$ where $\omega$ is an unknown exponent. 
We shall see that the algorithm is able to determine simultaneously and
self consistently both the best estimate for $\gamma_4$ and the exponent $\omega$. 

The plan of the paper is the following: in Sec.~\ref{Sec:Definition} we give some basic definition and review known facts
about the quartic coupling. We also  present our new results for the level expansion of $\gamma_4$.
In Sec.~\ref{Sec:BST} we discuss the properties of the BST algorithm and in Sec.~\ref{Sec:App} we apply it to the 
analysis of the LT series. Appendices are devoted to  technical details and tricks that we have found 
useful in the symbolic manipulation at high levels.

\section{Definition and level truncation results for $\gamma_4$}
\label{Sec:Definition}

Witten's open string field action is well known and reads
\be
S = \frac{1}{2\alpha'} \int \Psi\star Q\Psi + \frac{g}{3!} \int \Psi\star\Psi\star\Psi ,
\ee
where $Q$ is the BRST operator, $|\Psi\rangle$ is the ghost number 1 string field. Integration and $\star$ product
are well discussed in the literature and the reader can consult \cite{theses} for a review.
The gauge invariance of $S$ can be fixed by choosing the Feynman-Siegel gauge $b_0 |\Psi\rangle=0$.

As explained by Sen, the problem of tachyon condensation can be studied by restricting $\Psi$ to 
be a superposition with constant coefficients of open string  states with ghost number 1 
belonging to a certain universal space ${\cal H}$~\cite{Conjectures} defined, in the Feynman-Siegel gauge, by 
\be
{\cal H} = \mbox{Span}\left\{\prod_{i=1}^{N_m} L_{-k_i}\prod_{j=1}^{N_c-1} b_{-m_j}\prod_{l=1}^{N_c} c_{-n_l} |0\rangle,
\mbox{with}\ k_i\ge 2, m_j\ge 2, n_l=-1\ \mbox{or}\ n_l\ge 1\right\} ,
\ee
where $|0\rangle$ is the $SL(2,\mathbb R)$ invariant vacuum and $L_n$, $b_n$, $c_n$ are the modes of the matter energy-momentum tensor
and ghost fields $b$, $c$.
Due to a twist symmetry only states with even level $L\equiv L_0+1$ are relevant. Their number is summarized in Tab.~\ref{tab:11}. 
\TABLE{
\begin{tabular}{|| l cccccc|cccc || }
\hline
level     &$\le 10$& 12  & 14  & 16   & 18   & 20   & 22    & 24    & 26    & 28     \\
\hline
\# states & 170    & 231 & 496 & 1027 & 2060 & 4010 & 7611  & 14130 & 25697 & 45890  \\
total     & 170    & 401 & 897 & 1924 & 3984 & 7994 & 15605 & 29735 & 55432 & 101322 \\
\hline
\end{tabular}

\caption{Number of states at a given level in the universal subspace ${\cal H}$. The left part of the table corresponds to 
the most accurate existing calculation of the tachyon quartic coupling ($L\le 20$).}

\label{tab:11}
}

The evaluation of Witten's action can be separated in the calculation of the quadratic kinetic part
and the cubic interaction terms. The quadratic part in Feynman-Siegel gauge takes the simple form 
\be
S_2 = \frac{1}{2\alpha'} \int \Psi\star Q\Psi = \frac{1}{2\alpha'} \langle \Psi |\ c_0\ L_0\ | \Psi\rangle ,
\ee
where $\langle \Psi|$ is the BPZ conjugate of $|\Psi\rangle$. The evaluation of $S_2$
thus reduces to that of a long series of scalar products. These, in turn, can be computed separately in the matter and ghost sectors.
Selection rules can be exploited to know in advance which terms are necessarily vanishing. If we expand the field $|\Psi\rangle$ in 
components
\be
|\Psi\rangle = T\ c_1|0\rangle + \sum_s \varphi^s|s\rangle,\qquad |s\rangle\neq c_1|0\rangle,\ |s\rangle\in{\cal H} ,
\ee
($T$ being the tachyon field) the quadratic terms take the form
\be
S_2 = -\frac{1}{2\alpha'} T^2 + \frac{1}{2\alpha'}\sum_{ij} \mu_{ij}\ \varphi^i\varphi^j ,
\ee
where $\mu$ is a matrix with a well definite block structure. In particular $\mu$ connects only
fields with the same (matter + ghost) level.

The cubic terms, instead,  mix fields with different levels. For the purpose of computing the quartic tachyon 
coupling we need to consider only the $T^3$ term as well as the couplings $T^2\varphi$.
We denote the sum of such contributions by 
\be
\label{cubiccouplings}
S_3 = \kappa\ g\ T^3 + g\ T^2\sum_i c_i\varphi^i,
\ee
where $\kappa = 3^{7/2}/2^7$. The evaluation of $S_3$ can be split according to the level of $\varphi^i$. 
The full action is just a cubic polynomial in the space-time independent fields.

The (reduced) linear equations of motion of $\{\varphi\}$ obtained by differentiation of $S_2+S_3$ 
can be solved in terms of $T$. 
We stress that these are not the full equations of motion because $S_3$ contains only
a subset of the cubic couplings. Substituting the solution back in $S_2+S_3$ 
this procedure leads to the quartic tachyon potential
\be
\label{c4}
V(T) = -\frac{1}{2\alpha'}\ T^2 + \kappa\ g\ T^3-\frac 1 2\ g^2\ \alpha'\ \sum_{ij} (\mu^{-1})_{ij}c_i c_j\ T^4,
\ee
It is conventional to define a quartic coupling $\gamma_4$ independent on  $\alpha'$ and $g$ by writing 
the quartic term as $\kappa^2\alpha' g^2\gamma_4 T^4$. We shall adopt this notation and present our results in 
terms of $\gamma_4$.

Of course, the general construction of the tachyon potential requires to solve the full quadratic equations of motion of all fields with 
respect to $T$. The branch of the solution containing the unstable vacuum is the one with fields 
vanishing as $T\to 0$. Expanding in powers of $T$ we recover the perturbative
effective potential of the tachyon. For the quartic term, only the leading contributions are relevant and 
the very simple expression in Eq.~(\ref{c4}) can be used. This is a major simplification holding for the quartic term only.

To evaluate Eq.~(\ref{c4}) we need $\mu^{-1}$ and the cubic couplings $\{c_i\}$. 
The inversion of the mass matrices $\mu$ is  non trivial because we are interested in an exact result.
We must invert $\mu$ by symbolic manipulation and this is hard with the matrices at hand. To solve this technical problem
we have used a specific technique based on modular arithmetic and fully described in Appendix~\ref{App:Dixon}.

For the cubic interaction terms between states in ${\cal H}$ we have employed the conservation law technique discussed in~\cite{Conservation}.
The general term takes the form 
\be
\int \Psi\star\Psi\star\Psi = \sum_{i,j,k} \varphi^i\varphi^j\varphi^k\ 
\langle V_3| |i\rangle^{(1)}\otimes |j\rangle^{(2)}\otimes |k\rangle^{(3)} ,
\ee
where $|V_3\rangle$ is Witten's three string vertex\cite{OSFT}.
For the computation of $\gamma_4$ we just need the case that we write briefly as 
\ba
c_i &=& \langle V_3| |i\rangle^{(1)}\otimes c_1|0\rangle^{(2)}\otimes c_1|0\rangle^{(3)}  = \\
&=& \langle V_3| \prod_{a,b,c}L_{-n_a}b_{-m_b}c_{-l_c}|0\rangle^{(1)}\otimes c_1|0\rangle^{(2)}\otimes c_1|0\rangle^{(3)}. \nonumber
\ea
The action of the lowering operators on the dual vertex 
$\langle V_3|$ can be written in terms of positively moded operators acting on the three strings kets. 
The fast derivation of these conservation laws at arbitrary high level is 
described in Appendix~\ref{App:Conservation}.

The above procedure has been implemented in a {\em Mathematica} code. As a first check we have reproduced the various known contributions $\gamma_{4,L}$ from fields 
at fixed level $L$ for all $L\le 20$~\cite{Taylor00} reported in Tab.~\ref{tab:1} (the contributions from  odd level states  vanish).
\TABLE{
\begin{tabular}{|| c c c c  || }
\hline
&&&\\
L & $\gamma_{4,L}$ & L & $\gamma_{4,L}$ \\
&&&\\
\hline
&&&\\
2  & $\displaystyle-\frac{34}{27}$ & 
12 & $\displaystyle-\frac{3389771245837595}{183448998696332259}$ \\
&&&\\
4  & $\displaystyle-\frac{1399}{6561}$ & 14 & $\displaystyle-\frac{78109439317035676}{5853690776582965719}$ \\
&&&\\
6  & $\displaystyle-\frac{18016964}{215233605}$ & 16 & $\displaystyle-\frac{298001292970739836603}{29543127065508336986430}$ \\
&&&\\
8  & $\displaystyle-\frac{308423449}{6973568802}$ & 
18 & $\displaystyle-\frac{866910757208081799449068}{109838392116853446081848097}$ \\
&&&\\
10 & $\displaystyle-\frac{623826309310}{22876792454961}$ & 
20 & $\displaystyle-\frac{189246300010789017377006911}{29830815082559550621170156697}$  \\
&&&\\
\hline
\end{tabular}

\caption{Previously known LT results for the quartic tachyon coupling $\gamma_4$. The symbol $\gamma_{4,L}$ 
denotes the contribution to $\gamma_4$ from the fields with level $L$.}

\label{tab:1}
}

Then, we  computed the full exact contribution of states at levels $L=22$, $24$, $26$, $28$. 
The total number of states is thus $\sim 10^5$ to be compared with $\sim 8\cdot 10^3$ when only fields with level $L\le 20$ are 
considered. Our results are
\ba
\gamma_{4,22} &=& -\frac{1987896641921822231631291200}{381520424476945831628649898809} \simeq                        -0.005210,     \\ 
&&\nonumber \\
\gamma_{4,24} &=& -\frac{83592713383361636723926342390586}{19190858871614852276752718559991509} \simeq               -0.004356,          \\
&&\nonumber \\
\gamma_{4,26} &=& -\frac{6244227225924504591720556698980462}{1689629965870438080888011090607948075}\simeq            -0.003696,          \\
&&\nonumber \\
\gamma_{4,28} &=& -\frac{12670705380847526175723657846886589222}{3990838394187339929534246675572349035227}\simeq     -0.003175.
\ea

The resulting series for $\gamma_4$ has 14 terms and their sum is 
\be
-\frac{2^2\cdot 26863\cdot 36899\cdot 68230523\cdot 365310133\cdot 1823562056866280202643}
{3^{83}\cdot 5^2\cdot 11\cdot 13\cdot 17 \cdot 19\cdot 23} \simeq -1.70028 .
\ee
This figure can be compared with the numerical evaluation of the available closed form for $\gamma_4$. 
As we mentioned in the Introduction, the best result is $\gamma_4 = -1.742(1)$~\footnote{ 
The non negligible error is due to the singularities affecting the numerical quadratures required
for the calculation.}.
The discrepancy between this value and the partial sum of the LT series is 3.3\% at 
$L=20$ and 2.4\% at $L=28$. It is clear that the convergence is slow and seems to be roughly described by a leading 
correction vanishing as $1/L$.
This conjecture is supported in this specific case by the analysis of a different kind of level expansion,
described in~\cite{Diagrams} and based on an explicit description in terms of a reduced set of world-sheet 
diagrams.
It would be important to possess a reliable procedure to extrapolate the above table and predict
the asymptotic $L\to\infty$ result, possibly with some estimate of the extrapolation error.
Indeed, with a leading error of the form $1/L$ even a small error of about 1\% at level $L=10$ would require 
an unfeasible  calculation at level $L=100$ to reach a 0.1 \% precision.

\section{The extrapolation problem: polynomial and BST algorithms}
\label{Sec:BST}

Given the LT series, the simplest approach to determine the asymptotic value of $\gamma_4$ 
is that of performing a polynomial approximation in the variable $1/L$. In other words, given the 
values $\gamma_4(L) = \sum_{l=2}^L \gamma_{4,l}$ for a certain set of levels $L\in{\cal L}$ we choose an integer $N\le|{\cal L}|$ and solve the minimization 
problem
\be
\label{poly}
\sum_{L\in\cal L} \left| \gamma_4(L)-\sum_{n=0}^N a_n L^{-n}\right|^2 = \mbox{min} .
\ee
The resulting coefficient $a_0(N)$ is then chosen as an estimate of $\gamma_4$. In principle, better 
estimates can be obtained as $N$ is increased up to its maximum value $|\cal L|$. 

The problem with this (kind of) approach is that some assumption is implicit about the asymptotic
convergence of LT. In particular, the expansion in powers of $1/L$ is reasonable and supported by numerical 
data, but could nevertheless be misleading if LT convergence turned out to be dominated by leading
corrections of the form $1/L^\omega$ with an exponent close but not equal to one.

This effect can be seen if we can analyze a simple artificial example consisting of a short sequence of 10 pairs
$$
\left\{\left(h_n, f(h_n)\right)\right\}_{n=1, 2,\dots, 10},
$$
\be
\label{algebraic}
\{h_n\} = \left\{1, \frac 1 2 , \frac 1 3 , \dots , \frac{1}{10}\right\}, \qquad f(h) = 1+h^\alpha .
\ee
To determine numerically the limit $f(0)=1$, we can perform a polynomial fit as in Eq.~(\ref{poly}) with various $N$. For the exponent $\alpha$ 
we consider the following four values close to one: $\alpha = 0.8, 0.9, 1.1, 1.2$. 

In Fig.~\ref{fig:1} we plot
the estimate $a_0$ as a function of $N$ for the four considered $\alpha$. We see that as $N$ increases
the estimate of the asymptotic constant term improves, but at the largest possible $N$ there is still some 
non negligible disagreement.
\vskip 1.45cm
\FIGURE{\epsfig{file=fig1.eps,width=14cm} 
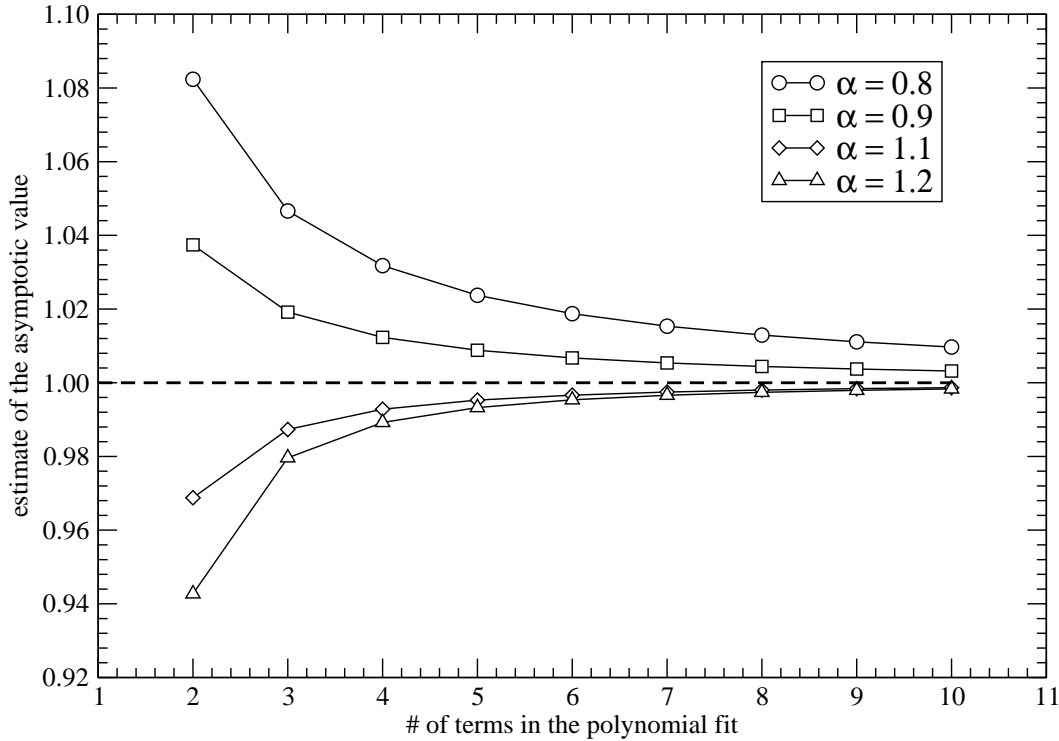
        \caption{Accuracy of the polynomial extrapolation.}
	\label{fig:1}}
It is difficult to obtain the exact value ({\em i.e.} $f(0)=1$) looking at the curves, especially when $\alpha<1$.
Even for $\alpha=0.9$ the error at the last point is of the order of the percent.

On general grounds, the extrapolation of the asymptotic value $f_\infty$ of a sequence $\{f_n\}$ 
is a relatively easy problem for linearly convergent sequences defined by the condition
\be
|\rho| < 1,\qquad\mbox{with}\ \rho = \lim_{n\to\infty}\frac{f_{n+1}-f_\infty}{f_n-f_\infty}  .
\ee 
In this case rigorous theorems~\cite{linear} assure the existence of algorithms able to 
accelerate the asymptotic convergence of any such sequence. Instead, when $\rho=1$,
the sequence is said to be logarithmically convergent and there is no general method that 
guarantees acceleration of \underline{all} sequences. The best that one can do is to consider a restricted class of 
logarithmically convergent sequences with a given asymptotic form such that an accelerating algorithms can be found.

An important example of this kind, relevant to our discussion, is that of 
sequences $f_n\sim f_\infty + a_1 n^{-\omega_1} + a_2 n^{-\omega_2} + \dots$
where $0<\omega_1<\omega_2<\dots$ are positive exponents. 
The convergence ratio is indeed $\rho=1$ and there is an accelerating procedure known in statistical mechanics 
as the BST algorithm~\cite{BST}. To describe it in full 
generality let us consider a function $f(h)$ admitting the following asymptotic expansion as $h\to 0$
\be
f(h) = f(0) + a_1 h^{\omega_1} + a_2 h^{\omega_2} + \cdots,\qquad 0 < \omega_1 < \omega_2 < \cdots ,
\ee
and suppose that we are given a set of pairs
\be
\{(h_n, f(h_n))\}_{0\le n < N},\qquad 0 < h_N < h_{N-1} < \cdots < h_0 .
\ee
The BST algorithm attempts to determine $f(0)$ by building a set of new sequences whose convergence is 
accelerated with respect to the initial one. To this aim, we build a grid of values $\{f_{n,m}\}_{0\le n < N, -1\le m < N}$ 
setting initially
\be
f_{n,-1}=0,\qquad f_{n,0} = f(h_n),
\ee
and defining the remaining values by the following iteration with respect to $m$:
\be
f_{n,m} = f_{n+1,m-1}+(f_{n+1,m-1}-f_{n,m-1})\left[\left(\frac{h_n}{h_{n+m}}\right)^\omega
\left(1-\frac{f_{n+1,m-1}-f_{n,m-1}}{f_{n+1,m-1}-f_{n+1,m-2}}\right)-1\right]^{-1} ,
\ee
where $\omega$ is a positive real number playing the role of a free parameter.

To apply the method, we compute the BST grid and, in particular, the final value $f_{0, N-1}(\omega)$ as functions of $\omega$. 
The best choice $\omega=\omega^*$ is the one that makes the last generated sequences ({\em i.e.} $f_{n,m}$ with $m=N-1$, $N-2$, $\dots$)
most flat. There is a certain amount of subjective judgement in this step, but usually, the optimal $\omega^*$
is identified in a quite narrow window. It is common to define a measure of the ``non-flatness'' of the last sequences
by constructing the sum of absolute differences between a certain set of points in the BST grid. 
In particular, we can define for a certain $K$ the border sum of differences
\be
\label{delta}
\delta_K(\omega) = 
\sum_{j=0}^{K-1} |f_{j,N-1-j}(\omega)-f_{j + 1, N-2-j}(\omega)| + \sum_{j=1}^K |f_{j, N-1-j}(\omega)-f_{j - 1, N-1-j}(\omega)| ,
\ee
and choose $\omega^*$ as the value that minimize $\delta_K(\omega)$. After this, the predicted limiting value 
is simply chosen to be $f_{0,N-1}(\omega^*)$.

As a final step, one would like to determine within reasonable bounds the prediction error. To this aim, the quantity $\delta_K(\omega^*)$ itself
can be used and, in practice, it works rather well. However, we strongly suggest to inspect the grid of $f_{n,m}$ values and
analyze the flatness of its subsequences in order to check this estimate.

We now repeat the analysis of our toy data set with the BST algorithm in order to appreciate the advantages
compared with the polynomial fit. We estimate the error using Eq.~(\ref{delta}) with $K=6$. 
The results are shown in Fig.~\ref{fig:2} where we plot the difference $\delta_6(\omega)$ and the 
prediction $f_{0,9}(\omega)$ as $\omega$ is varied in a wide range. 
For each considered $\alpha$ we can check that $\delta_6=0$ when $\omega=\alpha$. Also, as a further
check of the consistency of the algorithm, we can verify that $\delta_6$ vanishes at $\omega=\alpha/2$ too. 
The correct value $\omega=\alpha$ is thus selected automatically by the algorithm. Concerning the prediction, in the 
second part of Fig.~\ref{fig:2} we show that $f_{0,9}(\alpha) = 1$ within machine precision. These properties
are actually guaranteed for the BST algorithm that produces rigorously  a flat sequence after just two iterations when applied
to data following the simple law Eq.~(\ref{algebraic}). Fig.~\ref{fig:2} just gives a numerical check of 
these facts.

\vskip 1.3cm
\FIGURE{\epsfig{file=fig2.eps,width=12cm} 
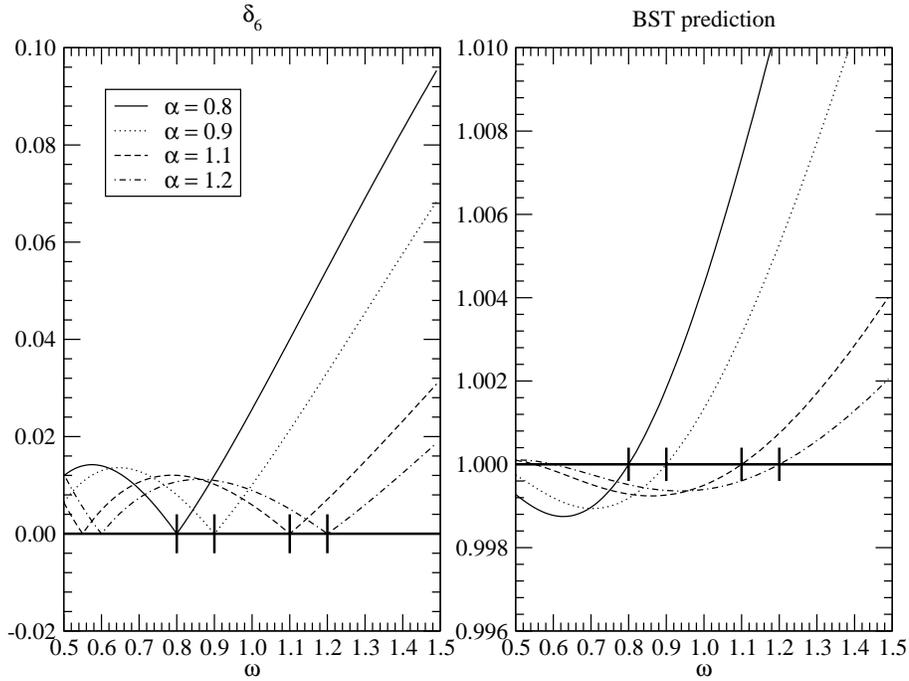
        \caption{Accuracy of the BST extrapolation.}
	\label{fig:2}}

The analysis of this simple example leads therefore to the following important conclusions.
First, the polynomial extrapolation 
can be apparently converging to stable values, but if the leading correction is $\sim 1/L^\omega$ with a
mismatch $\omega\neq 1$ it can  miss the correct result by a small but non zero amount.
Second, the BST algorithm finds consistently both the exact exponent of the leading correction and the asymptotic value.
It seems thus useful to examine the perturbative series obtained with level truncation by means of the BST
algorithm (or variants) to confirm and possibly refine the polynomial results.

\section{Determination of $\gamma_4$}
\label{Sec:App}

After this preliminary and general considerations, we consider the LT series for $\gamma_4$ and repeat the above analysis. 
In order to appreciate the information associated with the new computed terms we first perform 
the polynomial fit analysis on the first 10 terms and separately on the full extended series.
The results are summarized in Tab.~\ref{tab:3}; all the shown digits are meaningful since the LT terms are known exactly and 
the polynomial minimization can be performed with no numerical approximations.

Both sequences appear to converge well to stable values. In the first column (fit to 10 partial sums)
the difference between the last terms ($N=8,9$) is $\sim 10^{-6}$. In the second column (fit to 14 partial sums)
this difference (now between $N=12,13$) is instead reduced to $\sim 10^{-7}$. Therefore, the extended series affords
an estimate of $\gamma_4$ roughly one order of magnitude more precise in spite of the fact that the raw partial sum of the 
first 14 terms is only $\sim 28/20$ times closer to the predicted $-1.742(1)$ with respect to the one with 10 terms.
This means that the polynomial fit procedure is effective in accelerating the convergence.

What about the BST algorithm ? Is it able to accelerate even more the convergence rate ? 
To investigate this question, we apply it to the full sequence with the conservative choice $K=6$. 
\DOUBLETABLE{
\begin{tabular}{|| c c c || }
\hline
degree    & $L\le 20$ & $L\le 28$ \\
\hline
4         & -1.742020923  &  -1.742068107 \\	
5         & -1.742116782  &  -1.742143787 \\
6         & -1.742172515  &  -1.742181488 \\
7         & -1.742188837  &  -1.742192207 \\
8         & -1.742193921  &  -1.742195776 \\
9         & -1.742196399  &  -1.742197548 \\
10        & ---           &  -1.742198539 \\
11        & ---           &  -1.742199101 \\
12        & ---           &  -1.742199433 \\  
13        & ---           &  -1.742199638 \\
\hline
\end{tabular}
}
{
\begin{tabular}{|| c l l || }
\hline
L$\le$   & $\omega$ & $\gamma_4$ (BST) $\pm\delta_6$  \\
\hline
20 & 0.99896    &   -1.74221(6)   \\
22 & 0.99898    &   -1.74221(2)   \\
24 & 0.99965    &   -1.742202(3)  \\
26 & 0.9998     &   -1.7422009(9) \\
28 & 0.999875   &   -1.7422006(3) \\
\hline
\end{tabular}
}
{\label{tab:3}Results from the polynomial fit with data $L\le 20$ or $L\le 28$.}
{\label{tab:4}Results from the BST algorithm.}
The results are 
shown in Tab.~\ref{tab:4} where we collect the best exponent $\omega$ and the prediction for $\gamma_4$ when the
data at level $\le L$ are used.
From the table, we obtain the values $\gamma_4 = -1.7422006(3)$ and $|\omega-1| \lesssim 10^{-4}$.
The quality of the convergence can be deduced by a look at Tab.~\ref{tab:bst} where we show in the 5 columns
the last sequences generated by the algorithm in the analysis of the $L=28$ series.
\TABLE{
\begin{tabular}{|| cccccc || }
\hline
iter.:  & 9 & 10 & 11 & 12 & 13 \\
\hline
  &  1.7422001494 & 1.7422003937  &  1.7422005398   & {\bf 1.7422005287}  & {\bf 1.7422005986} \\
  &  1.7422003792 & 1.7422004666  &  {\bf 1.7422005294}   & {\bf 1.7422005175}  & - \\
  &  1.7422004648 & {\bf 1.7422004751}  &  {\bf 1.7422005189}   & -             & - \\
  &  {\bf 1.7422004966} & {\bf 1.7422005362}  &  -              & -             & - \\
  &  {\bf 1.7422005072} & -             &  -              & -             & - \\
\hline
\end{tabular}

\caption{Sample subset of the BST grid. The boldface figures are would be involved in the calculation of $\delta_4$.
We show only 10 digits, but the calculation has been done with much higher precision.}

\label{tab:bst}
}
The conclusion is that BST is able to achieve two important results: first, the value $\omega=1$
is strongly supported placing on firmer ground the underlying asumption made in the polynomial approach.
Second, $\gamma_4$ is estimated with an even better accuracy as can be seen looking at the 
differences between the boldface figures of Tab.~\ref{tab:bst}.

As a final comment, in order to assess the reliability of the prediction we can crosscheck it with another extrapolation 
technique as this is the safest procedure to determine the actual estimate error. A standard method for linearly convergent sequences and also for 
logarithmic ones with integer exponents (here $\omega\simeq 1$) is Levin $u$-transform~\cite{Levin}. 
We shall not discuss in details this algorithm,
but just quote the result that is obtained in the implementation  described in~\cite{LevinImp}. The application 
to the $L=28$ series gives $\gamma_4 = -1.7422009(6)$. The two results are completely compatible and we 
quote the BST central value with a very conservative error estimate as our final number 
\be
\gamma_4 = -1.7422006(9),\qquad |\omega-1| \lesssim 10^{-4}.
\ee

\section{Conclusions}

In this paper we have discussed the convergence of level truncation in open bosonic string field theory
focusing on the specific test case of the quartic tachyon coupling. We have pushed the computation 
of this quantity up to level $L=28$ and we have analyzed the resulting level expansion series by polynomial
fits and the  extrapolation BST algorithm. We have given numerical results supporting strongly 
the value  $\omega=1$ for the exponent of the leading corrections at finite $L$ of the form $\sim 1/L^\omega$.
Also, we have obtained a very precise determination of the coupling itself. The proposed methodology is
of general purpose and can be applied to any level truncation series. 

\acknowledgments

We thank C. Imbimbo and S. Giusto for conversations on tachyon condensation.

\appendix

\section{Modular arithmetic and Dixon's algorithm}
\label{App:Dixon}

In the calculation of the contributions to $\gamma_4$ from fields at a given level, 
we have to solve a linear problem associated to the mass matrix $\mu = \{\mu_{ij}\}$ 
to evaluate the last term in Eq.~(\ref{c4}). It is interesting to note that $\mu_{ij}\in\mathbb Z$. Indeed, the matrix elements of $\mu$ are factorized
\be
\mu_{ij} = \mu^{\rm matter}_{ij}\cdot\mu^{\rm ghost}_{ij}  .
\ee
The term $\mu^{\rm ghost}_{ij}$ can only be $0, \pm 1$ due to the ghost algebra. The term $\mu^{\rm matter}_{ij}$ is integer because it 
arises computing the scalar products of Virasoro descendents and the only dangerous (non integer) contribution can be 
the central charge term $c/12\ n(n^2-1)$. However, if $c=26$ it is evident that $\frac{13}{6}(n-1)n(n+1)\in\mathbb N$.

The components of the vector of known terms in the linear problem are the cubic couplings $c_i$ defined in Eq.~(\ref{cubiccouplings}).
Apart from a trivial $\sqrt{3}$ factor, these are rational numbers. Indeed, the factor $\sqrt 3$
arises in the basic 3 point ghost vertex $\langle V_3 | c_1|0\rangle \otimes c_1|0\rangle \otimes c_1|0\rangle$ and all the other manipulations do not introduce
irrational numbers. Multiplying by the largest denominator appearing in $\sqrt{3}\ c_i$ the problem can therefore 
be reduced to a totally integer problem $Ax=b$, with 
$A\in\mathbb Z^{n\times n}$ and $b\in\mathbb Z^n$. 

It is known that the exact symbolic solution of such a problem can be quite hard. In floating point arithmetics, the complexity of 
solving a linear problem is only polynomial (cubic in the matrix dimension). However, in infinite precision it is typical to 
generate very large integers in the intermediate steps of the solution and the complexity becomes exponential. The problem is not
trivial at the level we are working where the dimension of the relevant matrices can be as large as $\sim 700$ with 
integer entries with several tenths of digits.

A simple solution to this computational problem is afforded by Dixon's algorithm~\cite{Dixon}, a procedure arising in the field of modular 
p-adic arithmetic. 
Suppose we want to find the rational solution of 
\be
A x = b,\qquad A\in{\mathbb Z}^{n\times n}, \qquad b\in{\mathbb Z}^n .
\ee
We solve the problem in 3 steps. In the first step we choose a prime $p$ such that the following matrix exists:
\be
C = A^{-1}\  \mbox{mod}\ p .
\ee
The calculation of $C$ is not computationally expensive because all integers manipulated in the 
inversion of $A$ are bounded by $p$. In the second step we choose an integer $m$ and define
\be
b^{(0)} = b; \quad
x^{(i+1)} = C b^{(i)}\ \mbox{mod}\ p, \quad b^{(i+1)} = p^{-1}(b^{(i)}-Ax^{(i)}),\qquad 0\le i < p .
\ee
At the end of the iteration we set
\be
\bar x = \sum_{i=0}^{m-1} p^i x^{(i)}\in\mathbb Z^N .
\ee
It can be shown that $\bar x$ is a solution of the initial problem mod $p^m$. Finally, in the third step, 
we reconstruct the rational solution by solving the p-adic representation of $\bar x$. For each $1\le j \le n$ we define
\be
u^{(-1)} = p^m,\quad u^{(0)} = \bar x_j,\quad v^{(-1)} = 0,\quad v^{(0)} = 1 .
\ee
Denoting by $\mbox{floor}(x)$ the greatest integer less than or equal to $x$, we start with $i=0$ and iterate
\ba
\label{iteration}
\mbox{while} (u^{(i)} &>& p^{m/2}) \{ \\
q &=& \mbox{floor}(u^{(i-1)}/u^{(i)}) , \nonumber \\
u^{(i+1)} &=& u^{(i-1)}-q u^{(i)} ,\nonumber \\
v^{(i+1)} &=& v^{(i-1)}+q v^{(i)} ,\nonumber \\
i &\to& i+1, \nonumber \\
&\}& . \nonumber
\ea
At the end, we assign the $j$-th component of the proposed solution according to 
\be
x_j(p,m) = (-1)^i\frac{u^{(i)}}{v^{(i)}}
\ee
where $i$ takes the last value attained during the cycle in Eq.~(\ref{iteration}).
 The above procedure depends on the initial prime $p$ and the integer $m$
as emphasized in the notation.  
It can be shown that for any $p$, $x(p,m)$ equals the unique rational solution of the linear problem
problem as soon as $m$ exceeds a value that is in principle unknown, but that is assured to exist. Therefore,
in practice, one chooses a reasonably large $p$ (the larger $p$ the more expensive is the determination of $C$)
and than increases $m$ until $x(p,m)$ satisfies the initial system, a check that is not computationally
expensive.

\section{Conservation laws at high levels}
\label{App:Conservation}

As explained in~\cite{Conservation}, the evaluation of the cubic vertex among states belonging to ${\cal H}$
can be reduced to algorithmic rules that are fixed in terms of a set of 
meromorphic vector fields and quadratic differentials satisfying suitable conditions. Without repeating the details of the construction,
we just give a few simple formulas that permit a totally automatic determination of these objects with a negligible computational effort.
We hope that this Appendix will be useful for the reader interested in reproducing or extending this kind of calculation.

The discussion can be done separately for the matter and $c(z)$ ghost sectors. The conservation laws for the $b(z)$ ghost
are closely related to those in the matter sector. We define
\be
S(z) = \frac{z-\sqrt{3}}{1+z\sqrt{3}} ,
\ee
and expand in powers of $z$ the following maps
\ba
f_2(z) &=& \tan\left(\frac 2 3 \arctan z\right) = \frac 2 3 z -\frac{10}{81}z^3 + \frac{38}{729} z^5 - \frac{574}{19683} z^7 + \frac{30050}{1594323} z^9 + \cdots ,\\
f_3(z) &=& S(f_2) =  -{\sqrt{3}} + \frac{8}{3} z - \frac{16}{3\,{\sqrt{3}}} z^2 + \frac{248}{81} z^3 - \frac{416}{81\,{\sqrt{3}}}  z^4 + \cdots, \\
f_1(z) &=& S(f_3) = -f_3(-z) .
\ea

\subsection{Matter sector}

Following the notation of~\cite{Conservation}, the conservation laws for the Virasoro operators are determined by 
vector fields $v_n(z)$, $n\in\mathbb N$ of the form 
\be
v_n(z) = \frac{z^2-3}{z^{n-1}} V_n(z) ,
\ee
where $V_n$ is a polynomial with degree $[(n-1)/2]$ ($[x]$ denotes the integer part of $x$) and the following expansions
around $z=0$ hold
\be
\frac{v_n(f_2)}{f_2'} = \frac{1}{z^{n-1}} + 
\left\{\begin{array}{ll}
{\cal O}(z) & n \ \mbox{even} \\
{\cal O}(z^2) & n \ \mbox{odd}
\end{array}
\right. ,
\qquad 
\frac{v_n(f_{1,3})}{f_{1,3}'} = {\cal O}(z) .
\ee
In~\cite{Conservation}, the fields $v_n$ are built in sequence starting from $v_1$ and defining each of them in terms of suitable linear combination of the previous 
fields. This requires to fix the coefficients of the linear combination at each step. The whole procedure is not very expensive, but we think that it is 
interesting to give a compact formula for the result.
Solving to the desired order the constraints obeyed by $v_n$ we find:
\be
\frac{v_n(z)}{z^2-3} = \hat K\left(\frac{1}{z^2-3} \frac{1}{F'(z) F(z)^{n-1}}\right) ,
\ee
where $F(z) = \tan(3/2 \arctan z)$ and $\hat K f(z)$ is defined in terms of the Laurent expansion of $f(z)$:
\be
f(z) = \sum_{n=-N}^M c_n z^n \rightarrow \hat K(f(z)) = \sum_{n=-N}^0 c_n z^n .
\ee
The application of the algorithm determines very quickly all the vector fields $v_n$. To give an example, 
let us consider $n=10$. We have 
\be
\frac{1}{z^2-3} \frac{1}{F'(z)F(z)^9} =
\ee
$$
 = \frac{-1024}{177147\,z^9} + \frac{14336}{531441\,z^7} - \frac{99136}{1594323\,z^5} + \frac{461120}{4782969\,z^3} - \frac{1668820}{14348907\,z}
+{\cal O}(z),
$$
and thus acting with $\hat K$ and multiplying by $z^2-3$ we find
\be
V_{10}(z) = -\frac{4\, \left( 20736 - 96768\, z^2 + 223056\, z^4 - 345840\, 
        z^6 + 417205\, z^8 \right) }{14348907} .
\ee
The list of the first 9 vector fields is 
$$
V_1(z) = -\frac{2}{9}, \quad
V_2(z) = -\frac{4}{27}, \quad
V_3(z) = \frac{2\, \left( -4 + 7\, z^2 \right) }{81} 
$$
$$
V_4(z) = \frac{8\, \left( -6 + 13\, z^2 \right) }{729} , \quad
V_5(z) = -\frac{2\, \left( 48 - 124\, z^2 + 163\, z^4 \right) }{2187} 
$$
$$
V_6(z) = -\frac{4\, \left( 48 - 144\, z^2 + 217\, z^4 \right) }{6561} ,\quad 
V_7(z) = \frac{2\, \left( -576 + 1968\, z^2 - 3352\, z^4 + 3967\, z^6 \right) }{59049}
$$
$$
V_8(z) = \frac{16\, \left( -144 + 552\, z^2 - 1050\, z^4 + 1367\, z^6 \right) }{177147}
$$
$$
V_9(z) = -\frac{2\, \left( 2304 - 9792\, z^2 + 20592\, z^4 - 29336\, z^6 + 32969\, 
        z^8 \right) }{531441} .
$$

\subsection{Ghost sector}

In the ghost sector, the conservation laws for the modes of the $c(z)$ ghost field 
require the introduction of quadratic differentials $\phi_n(z)(dz)^2$ with general form 
\be
\phi_n(z) = \frac{\Phi_n(z)}{z^{n+2}(z^2-3)} ,
\ee
where $\Phi_n(z)$ is a polynomial of degree $2[n/2]$. The additional conditions that fix $\Phi_n$ are
\be
\phi_n(f_2) (f_2')^2 = \frac{1}{z^{n+2}} + 
\left\{\begin{array}{ll}
{\cal O}(1) & n \ \mbox{even} \\
{\cal O}(1/z) & n \ \mbox{odd}
\end{array}
\right. ,
\qquad 
\phi_n(f_{1,3})(f_{1,3}')^2 = {\cal O}(1/z) .
\ee
To find an explicit solution we compute 
\be
\hat K \left((z^2-3)\frac{F'(z)^2}{F(z)^{n+2}}\right) = 
\frac{c_{n+2}}{z^{n+2}} + \frac{c_{n+1}}{z^{n+1}} + \cdots + \frac{c_p}{z^p} ,
\ee
where $p=0(1)$ if $n$ is even(odd). The solution is then simply
\be
(z^2-3)\phi_n(z) = \frac{c_{n+2}}{z^{n+2}} + \frac{c_{n+1}}{z^{n+1}} + \cdots + \frac{c_{p+1}}{z^{p+1}} .
\ee
Again, we give an example. For $n=4$ we have 
\be
(z^2-3)\frac{F'(z)}{F(z)^6} = 
- \frac{16}{27 \ z^6} + \frac{16}{81\ z^4} + \frac{26}{81\ z^2} -\frac{200}{729} + {\cal O}(z^2) ,
\ee
and we obtain $\phi_4$ dividing the sum of the first 3 terms by $z^2-3$.
The first 10 cases are
$$
\Phi_0(z) = -3,\qquad
\Phi_1(z) = -2,\qquad
\Phi_2(z) = \frac{-2\, \left( 2 + z^2 \right) }{3},\qquad
\Phi_3(z) = \frac{-2\, \left( 12 + z^2 \right) }{27}
$$
$$
\Phi_4(z) = \frac{2\, \left( -24 + 8\, z^2 + 13\, z^4 \right) }{81},\qquad
\Phi_5(z) = \frac{2\, \left( -48 + 36\, z^2 + 17\, z^4 \right) }{243}
$$
$$
\Phi_6(z) = \frac{-2\, \left( 288 - 336\, z^2 + 2\, z^4 + 217\, z^6 \right) }{2187}
$$
$$
\Phi_7(z) = \frac{-2\, \left( 576 - 912\, z^2 + 312\, z^4 + 391\, z^6 \right) }{6561}
$$
$$
\Phi_8(z) = \frac{2\, \left( -1152 + 2304\, z^2 - 1440\, z^4 - 416\, z^6 + 1367\, 
        z^8 \right) }{19683}
$$
$$
\Phi_9(z) = \frac{2\, \left( -20736 + 50112\, z^2 - 44208\, z^4 + 5640\, z^6 + 25703\, 
        z^8 \right) }{531441}
$$
$$
\Phi_{10}(z) = \frac{-2\, \left( 41472 - 117504\, z^2 + 132192\, z^4 - 53616\, z^6 - 40622\, 
        z^8 + 83441\, z^{10} \right) }{1594323} .
$$

\end{document}